\documentstyle[bo99,epsfig]{article}

\title{Discovery of hard X-ray emission from\\
Type II bursts of the Rapid Burster
}
\author{F. Frontera$^{1,2}$, N. Masetti$^1$, M. Orlandini$^1$, L. Amati$^1$,
E. Palazzi$^1$, D. Dal Fiume$^1$, S. Del Sordo$^3$, G. Cusumano$^3$, A.N.
Parmar$^4$, G. Pareschi$^5$, I. Lapidus$^6$ and L. Stella$^7$}

\affil{$^1$Istituto Te.S.R.E., CNR, via Gobetti 101, I-40129 Bologna,
Italy\\
$^2$Dipartimento di Fisica, Universit\`a di Ferrara, via Paradiso 12,
I-44100 Ferrara, Italy\\
$^3$I.F.C.A.I., CNR, via Ugo La Malfa 153, I-90146 Palermo, Italy\\
$^4$SSD, ESA/ESTEC, Postbus 299, 2200 AG Noordwijk, The Netherlands\\
$^5$Osservatorio Astronomico di Brera, Via Bianchi 46, I-23807 Merate,
Italy\\
$^6$McKinsey \& Co. Inc., London, UK\\
$^7$Osservatorio Astronomico di Roma, via Frascati 33, I-00040
Monteporzio Catone, Italy}

\begin{document}

\maketitle

\begin{abstract} We report on results of {\it BeppoSAX} Target Of
Opportunity (TOO) observations of the source MXB 1730-335, also called the
Rapid Burster (RB), made during its outburst of February--March 1998. We
monitored the evolution of the spectral properties of the RB from the
outburst decay to quiescence.  During the first TOO, the
X--ray light curve of the RB showed many Type II bursts and its
broadband (1-100 keV) spectrum was acceptably fit with a two blackbody
plus power law model. Moreover, to our knowledge, this is the
first time that this source is detected beyond 30 keV. 

\end{abstract}

\section{Observations}

Four Target Of Opportunity (TOO) observations were performed with
{\it BeppoSAX} (Boella et al. 1997a) on the Rapid Burster (=MXB 1730--335;
hereafter RB) during the activity state which started on January 28, 1998
(Fox et al. 1998).
These TOOs spanned over one month (from February 18 to March 18) and
caught the object in four different snapshots, from the post--maximum
decay to the quiescent state. Figure 1, left panel, shows the ASM
light curve of the {\it Rossi-XTE} satellite with superimposed the times
of the four {\it BeppoSAX} observations.
Here we report on RB data from three of the four instruments mounted on
{\it BeppoSAX}: LECS (0.1-10 keV; Parmar et al. 1997), MECS (1.5-10
keV; Boella et al. 1997b, and PDS (15-300 keV; Frontera et al. 1997). 
For the PDS the default rocking collimator law was modified by offsetting 
the RB by 40$^{'}$ from the center of the field of view in order to reduce
as much as possible the contamination from a nearby variable X--ray
source, GX 354-0 (=4U 1728-34), located at about 30$^{'}$ from the RB.
Unfortunately, due to failure of the rocking law setup program, the
collimator did not move as requested during TOO2 and TOO3; so, we have
only LECS and MECS data for these two observations.

In this paper we report on preliminary results of these observations.
Definitive results along with their implications will be the subject of
another paper (Masetti et al. 2000).
In the following, for the luminosity estimates we will assume that the
RB lies at a distance $d$ = 8 kpc (Ortolani et al. 1996).

\begin{figure}
\begin{center}
\epsfig{file=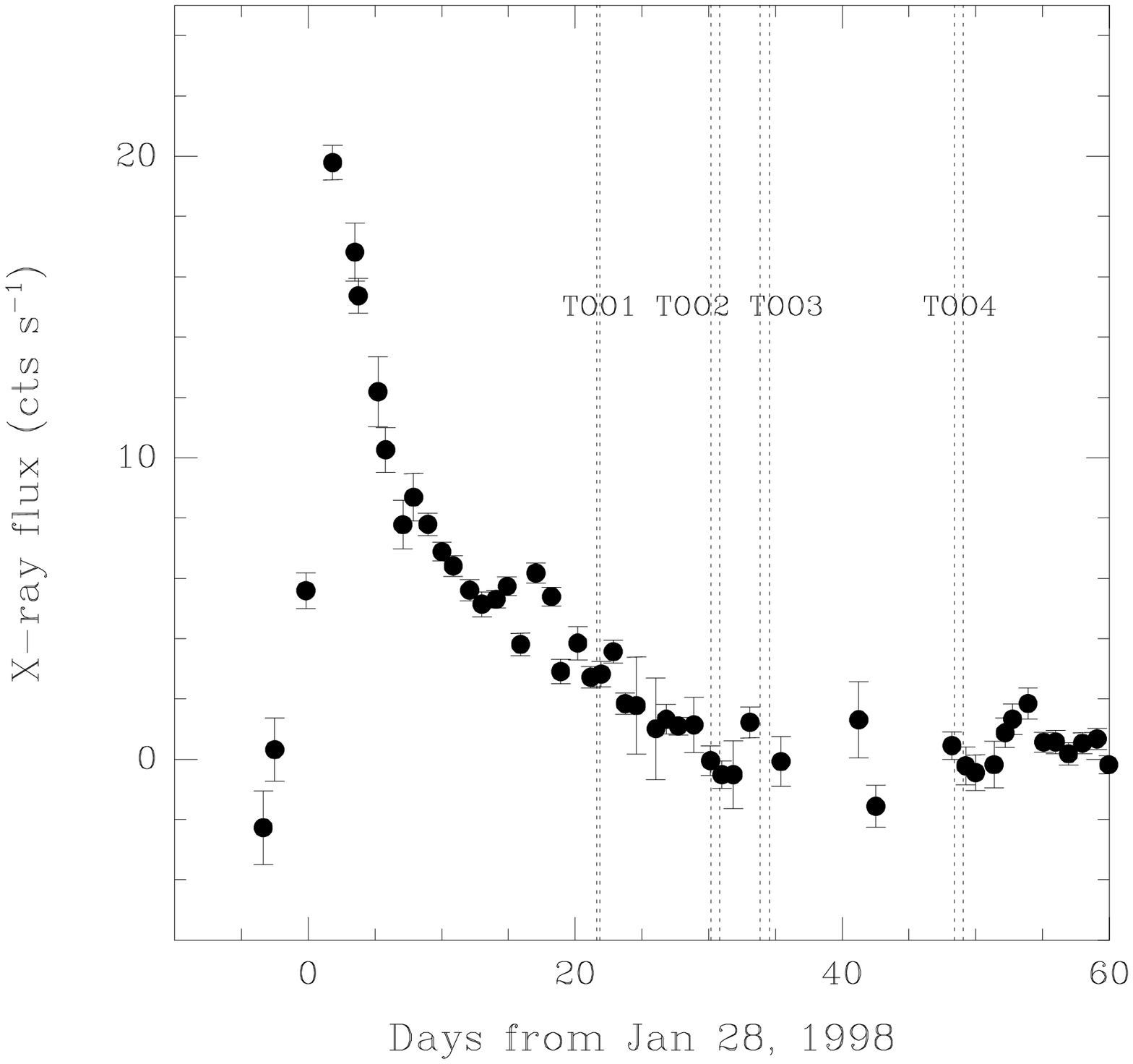, height=6.5cm}
\epsfig{file=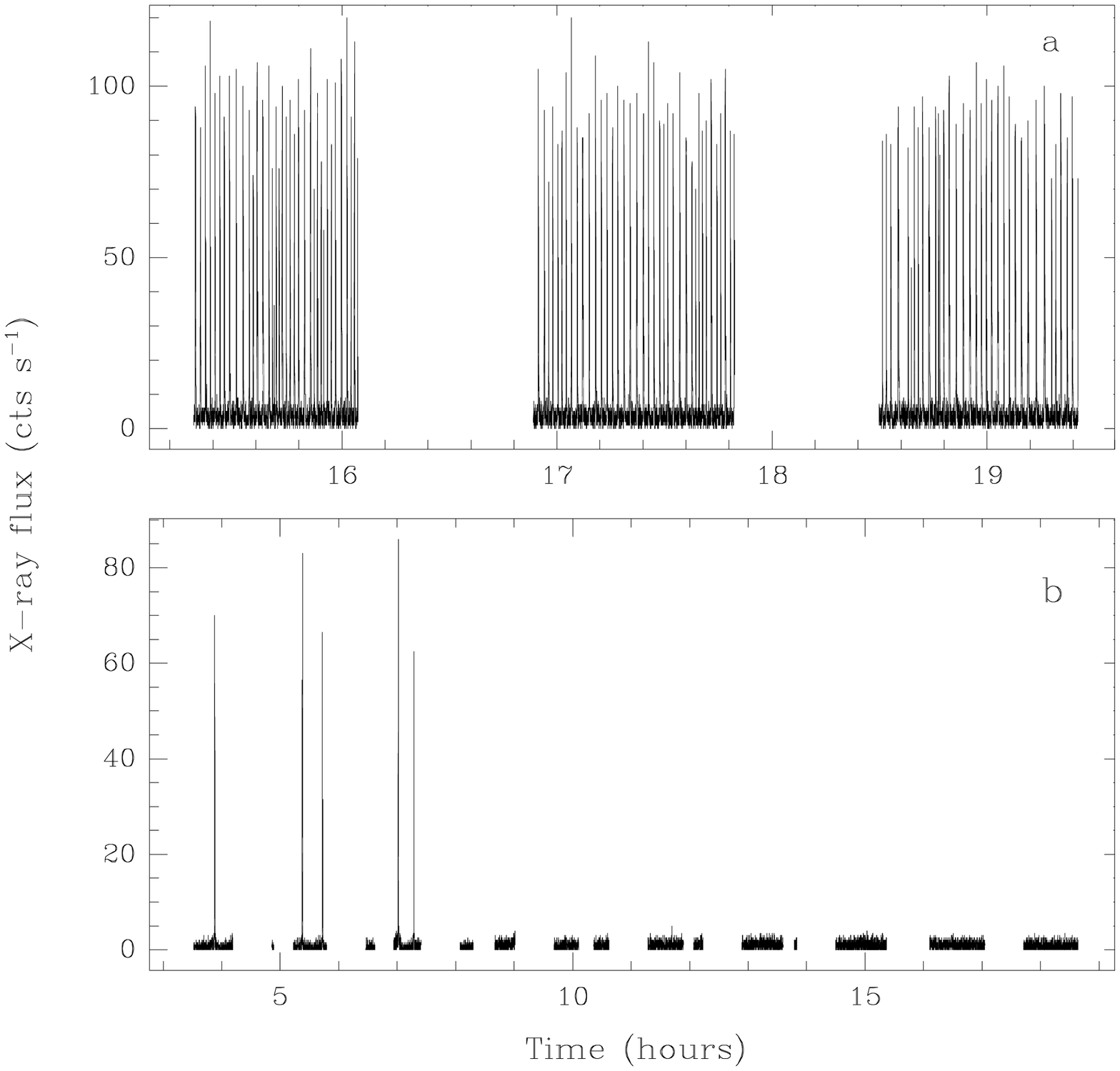, height=6.0cm}
\end{center}
\vspace{-1.4cm}
\caption[]{{\it (Left panel)} {\it XTE} ASM light curve of the
February-March 1998 activity phase of the RB. The vertical dashed lines
indicate the four {\it BeppoSAX} TOOs. {\it (Right panel)} ({\bf a}) MECS
light curve of TOO1. Times are expressed in
hours from 0 UT of February 18, 1998; ({\bf b}) MECS light curve of
TOO2. Times are expressed in hours from 0 UT of February 27, 1998.
}
\end{figure}

\section{Spectral analysis and temporal evolution of the RB}

During TOO1, the RB was in a strong state of bursting activity.
The 2-10~keV light curve obtained with MECS (see Fig. 1, right panel,
part {\it a}) showed 113 Type
II X--ray bursts during 9457 seconds of good observational data.
Evidence of Type II bursts was also observed in the 0.1-2~keV data
obtained with LECS. 
We divided the MECS TOO1 data into two subsets: persistent emission
(PE; below 5 counts s$^{-1}$) and bursting emission (BE; above 5 counts
s$^{-1}$).

The MECS PE and BE spectra could be well fit  
with a photoelectrically absorbed two-component blackbody (2BB); these BB
components may originate from the neutron star (NS) surface, a
boundary layer between the NS and the inner edge of the accretion disk, or
the inner region of the disk itself.
The same model was used for the RB by Guerriero et al. (1998) who found
values consistent with ours. 
In Table 1 we report the best-fit parameters along with their 90\%
confidence errors.
The temperature values of the two BB components were slightly 
higher during the PE than during the BE, while their
luminosities were much higher (by a factor 20 for the cooler BB and 60 for 
the hotter BB) during the BE than during the PE. We also remark that
during the BE the hotter BB component was brighter (by more than a
factor 3) than the cooler BB, while during the PE they had similar
luminosities.
This implies that the BE influences more the higher temperature component
than the other one.
If the BE is due to spasmodic accretion onto the compact object, the
higher temperature component should be the one coming from the NS surface.  


\begin{table}
\caption[]{Best-fit parameters for the TOO spectra. The reported
uncertainties are given at 90\% confidence level for a single parameter
of interest.}
\smallskip
{\footnotesize
\begin{tabular}{c|ccc|c|c}
\hline
\noalign{\smallskip}
 & \multicolumn{3}{c|}{TOO1} & TOO2 & TOO3 \\
\noalign{\smallskip}
\hline
\noalign{\smallskip}
Model & 1-10 keV & 1-10 keV & 1-100 keV & 0.5-10 keV & 0.5-10 keV \\
 & BE & PE & BE $-$ PE & (mainly PE) & PE \\
\noalign{\smallskip}
\hline
\noalign{\smallskip}
${\chi^{2}_{\nu}}$~(dof) & 1.17~(395)& 1.07~(203)& 1.17~(223)& 0.97~(113)&
1.07~(55)\\
\multicolumn{1}{l|}{Wabs(2BB):} & & & & & \\ 
$N_{\rm H}$ ($\times$10$^{22}$ cm$^{-2}$) & $3.5\pm0.5$ & $1.5\pm0.3$ &
$4.9^{+1.7}_{-1.1}$ & $1.6\pm0.3$ & $1.1^{+0.4}_{-0.3}$ \\
$kT_1$ (keV) & $0.43\pm0.04$ & $0.63\pm0.06$ &
$0.32^{+0.23}_{-0.12}$ & $0.65\pm0.07$ & $0.64\pm0.09$ \\
$L^{\rm BB}_1$$^{(a)}$ & $15^{+6}_{-5}$& $0.85^{+0.08}_{-0.06}$ &
$10^{+22}_{-10}$ & $0.32^{+0.04}_{-0.03}$ & $0.06\pm0.01$ \\
$kT_2$ (keV)& $1.64\pm0.03$ & $1.72^{+0.19}_{-0.17}$ &
$1.67\pm0.04$ & $1.78^{+0.10}_{-0.08}$ & $2.1^{+0.4}_{-0.2}$ \\
$L^{\rm BB}_2$$^{(a)}$ & $47\pm0.8$ & $0.74^{+0.06}_{-0.05}$ &
$40\pm2$ & $0.76\pm0.03$ & $0.106^{+0.011}_{-0.008}$ \\
\multicolumn{1}{l|}{+ power law:} & & & & & \\
$\Gamma$ & & & $3.1^{+0.3}_{-0.4}$ & & \\
$K$$^{(b)}$ & & & $3.3^{+4.5}_{-2.5}$ & & \\
\multicolumn{1}{l|}{+ Fe emission line:} & & & & & \\   
$E_{\rm l}$ (keV) & & & $6.5\pm0.2$ & & \\
EW (eV) & & & $100^{+80}_{-60}$ & & \\
FWHM (keV) & & & $0.4^{+0.3}_{-0.4}$ & & \\
$I_{\rm l}$$^{(c)}$ & & & $7^{+5}_{-4}$ & & \\
\noalign{\smallskip}
\hline
\noalign{\smallskip}
\multicolumn{6}{l}{$^{(a)}$: in units of 10$^{36}$ erg s$^{-1}$} \\
\multicolumn{6}{l}{$^{(b)}$: in units of photons keV$^{-1}$ cm$^{-2}$
s$^{-1}$ at 1 keV} \\
\multicolumn{6}{l}{$^{(c)}$: in units of 10$^{-3}$ photons cm$^{-2}$
s$^{-1}$} \\
\end{tabular}
}
\end{table}


The source was also visible in the hard X--ray (15-100 keV) energy range.
However the statistics of the PDS light curve was much lower and did not
allow distinguishing the Type II bursts. In order to construct the BE
spectrum we used the time intervals in which the bursts were observed with
MECS. Also, we could not derive the correct 15-100 keV flux and spectrum
of both BE and PE given the residual source contamination by GX 354-0.
Thus, in order to overcome this problem, we used as background level
for the 1-100 keV BE spectrum the total count rate level measured during
the PE time intervals.
The combined LECS+MECS+PDS PE-subtracted bursting spectrum, shown in Fig.
2, was no longer fit with a 2BB model. By adding a power law component
we obtained an acceptable fit (see Table 1). 
The further addition of a Fe K emission line at 6.5 keV slightly improved
the fit, with parameter values found for this line in general agreement
with the findings by Stella et al. (1988) for Type II bursts.

\smallskip
During TOO2 the RB drastically
reduced its bursting activity, and the bursts were concentrated at the
beginning of this TOO (Fig. 1, right panel, part {\it b}). Also, the
emission intensity level decreased. 
The best--fit model was a photoelectrically absorbed 2BB model (Table 1).
No evidence of a Fe emission line was present.

During TOO3 the object further reduced its bursting activity, and no 
Type II bursts were seen throughout the observation. 
The best--fit model spectrum was still an absorbed 2BB (Table 1). 
As in the case of TOO2, no iron emission line at 6.5 keV was found.

The RB was instead no longer visible in MECS/LECS images during TOO4. 
Stray light from GX354-0 prevented us to get a deep observation of the
source. The 3$\sigma$ upper limit to the RB X--ray emission was 
1.5$\times$10$^{-12}$ erg cm$^{-2}$ s$^{-1}$ in the 2-10 keV energy band.

\begin{figure}
\begin{center}
\epsfig{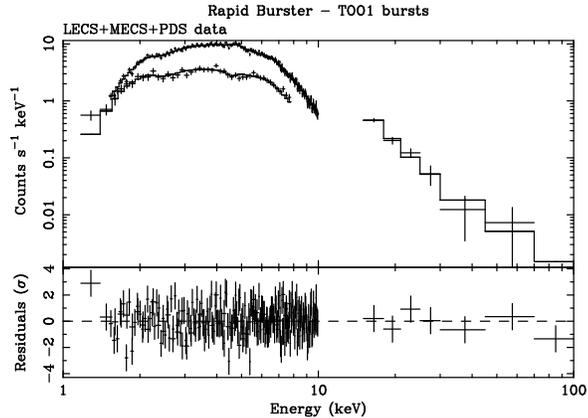}
\end{center}
\vspace{-.5cm}
\caption[]{TOO1 LECS+MECS+PDS X--ray spectrum of the
total PE-subtracted BE. The fit corresponds to an absorbed two-component
blackbody plus power-law model plus an iron line at 6.5 keV.}
\end{figure}

\end{document}